\documentclass[10pt,sigplan,letterpaper,twocolumn]{article}
\usepackage[10pt,nocopyright]{sigmin}

\usepackage{listings}
\usepackage{color}
\usepackage{epsfig,makecell}    
\usepackage{ifpdf}
\ifpdf
  \usepackage{epstopdf}
\fi                                  
\usepackage[square,comma,numbers,sort&compress]{natbib}          
\usepackage[hyphens]{url}                                         
\usepackage[breaklinks,colorlinks]{hyperref}                      
\usepackage[usenames,dvipsnames]{xcolor}                          
\hypersetup{citecolor=blue,linkcolor=blue}                        
\usepackage{amsmath,amsopn,amssymb,amsthm}                        
\usepackage{thmtools}
\usepackage[toc,page]{appendix}
\usepackage{hyperref}
\usepackage{subfigure}
\usepackage{endnotes,microtype,xspace,fancyvrb,multirow} 
\usepackage{booktabs}                                             
\usepackage{array,underscore,relsize}                             
\usepackage[T1]{fontenc}                                          
\usepackage{times}                                                
\usepackage{fancyhdr,lastpage}                                    
\usepackage{enumitem}                                             
\usepackage[labelfont=bf,font=normal,skip=1pt]{caption}           
\usepackage[export]{adjustbox}                                    
\usepackage{breakurl}                                             
\usepackage{pifont}                                               
\usepackage{tablefootnote}                                        
\usepackage{graphicx}
\usepackage[linesnumbered,vlined,ruled,commentsnumbered]{algorithm2e}
\DontPrintSemicolon
\newcommand{\ra}[1]{\renewcommand{\arraystretch}{#1}}
\SetCommentSty{mycommfont}

\usepackage{amssymb}
\usepackage[normalem]{ulem}
\usepackage[hang,flushmargin]{footmisc}
\usepackage{textcomp}
\usepackage{graphicx}

\usepackage[compact]{titlesec}
\titleformat*{\section}{\large\bfseries}
\titleformat*{\subsection}{\normalsize\bfseries}
\titleformat*{\subsubsection}{\normalsize\bfseries}
\titlespacing*{\section}{0pt}{*2.7}{*1}
\titlespacing*{\subsection}{0pt}{*1.5}{*0.8}

\definecolor{mygreen}{rgb}{0,0.6,0}  
\definecolor{mygray}{rgb}{0.5,0.5,0.5}  
\definecolor{mymauve}{rgb}{0.58,0,0.82}

\hypersetup{
  colorlinks,
  linkcolor={red!50!black},
  citecolor={blue!50!black},
  urlcolor={blue!80!black}
}

\newcommand{\sys}{{\textsc{Picker}}}

\newcommand{\us}{{{$\upmu$}s}}

\newcommand{\pidem}{{conditionally-idempotent}}
\newcommand{\pidemshort}{{cond-idempotent}}
\newcommand{\pidemcap}{{Conditionally-idempotent}}

\newcommand{\noidentstitle}[1]{\vspace{0.5ex}\noindent{\bf #1}}
\newcommand{\stitle}[1]{\vspace{1.ex}\noindent{\bf #1}}
\newcommand{\etitle}[1]{\vspace{0.8ex}\noindent{\em\underline{#1}}}

\newcommand{\code}[1]{\texttt{#1}}

\newcommand{\squishlist}{
  \begin{list}{$\bullet$}{
    \setlength{\itemsep}{0pt}
    \setlength{\parsep}{3pt}
    \setlength{\topsep}{3pt}
    \setlength{\partopsep}{0pt}
    \setlength{\leftmargin}{1.5em}
    \setlength{\labelwidth}{1em}
    \setlength{\labelsep}{0.5em}
  }
}

\newcommand{\squishend}{
  \end{list}
}

\usepackage{tikz}

\newcommand*\emptycirc[1][1ex]{\tikz\draw (0,0) circle (#1);} 
\newcommand*\halfcirc[1][1ex]{%
  \begin{tikzpicture}
  \draw[fill] (0,0)-- (90:#1) arc (90:270:#1) -- cycle ;
  \draw (0,0) circle (#1);
  \end{tikzpicture}}
\newcommand*\fullcirc[1][1ex]{%
  \begin{tikzpicture}
  \draw (0,0) circle (#1);
  \draw[fill] (0,0) circle (#1);
  \end{tikzpicture}}

\usepackage{balance}
\usepackage{upgreek}

\begin{document}

\title{\Large \bf Microsecond-scale Dynamic Validation of Idempotency for GPU Kernels}

\author{
    Mingcong Han,\;
    Weihang Shen,\;
    Guanwen Peng,\;
    Rong Chen\thanks{Rong Chen is the corresponding author (\url{rongchen@sjtu.edu.cn}).},\;
    and\; Haibo Chen\;\\[5pt]
    \normalsize{Institute of Parallel and Distributed Systems, Shanghai Jiao Tong University}
}

\date{}
\maketitle

\frenchspacing

\begin{abstract}

We discovered that a GPU kernel can have both idempotent and non-idempotent instances
depending on the input. 
These kernels, called {\pidem}, are prevalent in real-world 
GPU applications (490 out of 547 from six applications).
Consequently, prior work that 
classifies GPU kernels as either idempotent or non-idempotent 
can severely compromise the correctness or efficiency of idempotence-based systems. 
This paper presents {\sys}, the first system for instance-level idempotency validation.
{\sys} dynamically validates the idempotency of GPU kernel instances before their execution, 
by utilizing their launch arguments.
Several optimizations are proposed to significantly reduce validation latency to microsecond-scale.
Evaluations using representative GPU applications (547 kernels and 18,217 instances in total) 
show that {\sys} can identify idempotent instances with no false positives
and a false-negative rate of 18.54\%, and can complete the validation within 5\,{\us} for all
instances. 
Furthermore, by integrating {\sys}, 
a fault-tolerant system can reduce the checkpoint cost to less than 4\% and 
a scheduling system can reduce the preemption latency by 84.2\%.

\end{abstract}

\section{Introduction}
Idempotent GPU kernels promise to produce the same output upon re-execution, 
without causing any side effects, no matter how many times it is interrupted
at any point.
Such idempotence property has been shown to be effective in overcoming 
the performance bottlenecks of many aspects of GPU systems, 
including fault tolerance~\cite{Leng2020AsymmetricRE,Zhang2020FaulttolerantAT}, 
task preemption~\cite{han2022reef,Park2015ChimeraCP,Lee2021IdempotenceBasedPG},
and memory persistence~\cite{DBLP:conf/iiswc/YudhaKZS20,Lin2019ExploringMP}.
For instance, prior work shows that idempotence-based optimizations can significantly reduce
the checkpointing overheads of fault-tolerant systems to less than 1\%~\cite{Leng2020AsymmetricRE}.


However, given the presence of non-idempotent GPU kernels,
idempotence-based optimizations require \textbf{prior} knowledge of kernel idempotency 
(i.e., knowing before execution).
This idempotency information must be \textbf{correct}, ensuring that no non-idempotent kernels
are incorrectly identified as idempotent (i.e., no false positives);
otherwise, the system may erroneously re-execute a non-idempotent kernel from an inconsistent state.
Moreover, it is also important for the idempotency information to be \textbf{efficient}, 
identifying as many idempotent functions as possible (i.e., fewer false negatives), 
thereby improving the system efficiency by leveraging the benefits of idempotency.
Existing systems, however, determine the idempotency of a GPU kernel by statically 
analyzing either the source code~\cite{Lee2021IdempotenceBasedPG,Park2015ChimeraCP,Leng2020AsymmetricRE}
or historical traces~\cite{han2022reef}.

\begin{figure}[t]
\vspace{1mm}
\begin{minipage}{1.\linewidth}
\centering\includegraphics[width=1.\linewidth]{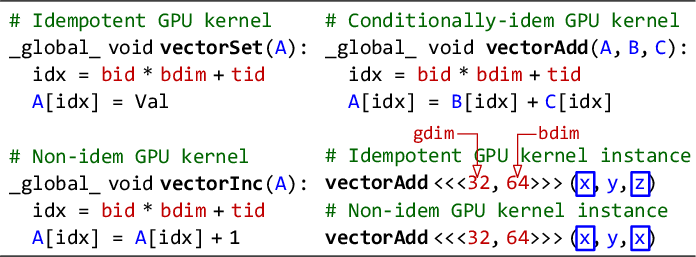}
\end{minipage} \\[8pt]
\begin{minipage}{1.\linewidth}
\caption{\emph{\small{
Simplified version of an idempotent GPU kernel, a non-idempotent GPU kernel,
and a {\pidem} GPU kernel with its two instances. 
\textup{\code{bid}}, \textup{\code{tid}}, \textup{\code{gdim}} and \textup{\code{bdim}} 
stand for the index and the dimension of a block and a thread, respectively.}}}
\label{fig:vectoradd}
\end{minipage}  \\[-15pt]
\end{figure}

\stitle{New finding.} 
We discovered that a GPU kernel can have both idempotent and non-idempotent instances,
where an instance refers to the invocation of a GPU kernel with a specific input state and arguments.
These \emph{{\pidem}} (abbreviated as {\pidemshort}) GPU kernels are common in real-world
GPU applications.
For instance, 490 out of 547 kernels 
we have studied are {\pidemshort}, including all kernels generated by TVM 
(see Table~\ref{tbl:kernels}).\footnote{\footnotesize{We also discovered four
{\pidemshort} GPU kernels in their public application traces from Rodinia~\cite{Che2009RodiniaAB}
and PyTorch~\cite{PyTorch} (see \S\ref{sec:study} for details).}}
This is because GPU kernels usually have multiple pointer arguments that are passed by 
the host program. 
If the memory used for input and output overlaps,
the GPU kernel instance is likely to be non-idempotent as the input data may be overwritten
during execution.
For example, Fig.~\ref{fig:vectoradd} shows a {\pidemshort} GPU kernel (\code{vectorAdd})
and its two kinds of instances.

However, prior work statically classifies a GPU kernel as either fully idempotent or
non-idempotent at the \textbf{kernel level}~\cite{Lee2021IdempotenceBasedPG,Leng2020AsymmetricRE,Park2015ChimeraCP,han2022reef},
and assumes that the idempotency of all instances of a kernel are the same.
This approach overlooks the existence and prevalence of {\pidemshort} GPU kernels.
Consequently, classifying {\pidemshort} kernels as purely idempotent compromises
the correctness of idempotence-based systems.
Conversely, treating {\pidemshort} kernels as purely non-idempotent
severely sacrifices the effectiveness of idempotence-based optimizations,
reducing their applicability by around 90\% (see \S\ref{sec:study}). 

Our study reveals that, despite most GPU kernels being {\pidemshort}, nearly 80\% of instances 
are idempotent according to public traces 
(see Table~\ref{tbl:instances}).
This finding motivates us to 
identify idempotency at the \textbf{instance level},
as it can satisfy both correctness and efficiency.
However, accurately determining the idempotency of an instance is only feasible after its
execution, which contradicts the requirement of idempotence-based systems to know it prior
to execution.

\stitle{Key insight.}
We found that in most cases, it is sufficient to identify the idempotency of a 
GPU kernel instance by examining the value of its launch arguments, which are known
without execution. This opens an opportunity for dynamic validation of instance-level
idempotency prior to execution.

\stitle{Our approach.}
To this end, we present {\sys}, the first system that validates instance-level 
idempotency for GPU kernels.
{\sys} dynamically intercepts the GPU kernel launch calls at runtime, and
validates the idempotency of the corresponding instance before it is submitted 
to the idempotence-based system.

Specifically, {\sys} first transforms the problem of instance-level idempotency validation 
into the problem of determining the memory addresses accessed by the instance,
which can be validated before the instance is executed. 
{\sys} considers an instance as idempotent if it can validate that
there are no memory addresses that may be both read and written, for every byte
that could be accessed by the instance.
Once an address has the potential to be both read and written, it is classified
as non-idempotent.

\stitle{Challenge: microsecond-scale dynamic validation.}
In contrast to the kernel-level approach which completes all analysis offline,
dynamic validation on a GPU instance imposes runtime overhead 
to the critical path of the GPU kernel launch. 
The execution of a GPU kernel instance usually begins within several microseconds 
after it is launched.
Therefore, {\sys} should also complete the validation within a few microseconds,
allowing it to validate only a few dozen memory addresses.
However, the complexity arises from the fact that a GPU kernel instance typically
employs thousands of GPU threads collectively accessing a tremendous amount of memory
addresses. It is extremely challenging to examine such a large quantity of 
memory addresses (up to billions) within such a short timeframe (a few microseconds).

To tackle this challenge, {\sys} represents the memory addresses accessed by all GPU 
threads at the same load/store instruction using a \textit{range}, based on the 
observation that these addresses are often \textit{consecutive}.
By estimating the upper and lower bounds of this range, it avoids the need
to enumerate every possible address accessed by all GPU threads.
To accelerate the calculation of the bounds of the range, 
{\sys} capitalizes on the observed \textit{monotonicity} of the GPU thread IDs
and the accessed addresses.
Specifically, it means that larger GPU thread IDs correspond to higher accessed addresses
for the same load/store instruction.
By leveraging this monotonicity, {\sys} simplifies the calculation of a range
by only considering the GPU threads with the minimum and maximum IDs.
With these optimizations, {\sys} can reduce the validation time from
milliseconds to microseconds, minimizing the overhead to the application.

{\sys} is designed to directly analyze the assembly code of GPU kernels (i.e., SASS code for NVIDIA 
GPU~\cite{NvidiaISA,Hayes2019DecodingCB}), 
making it applicable to production-level GPU applications 
with closed-source or dynamically generated kernels.
We evaluated the efficacy of {\sys} in validating idempotency using
applications from well-known GPU benchmarks (i.e., Rodinia~\cite{Che2009RodiniaAB} 
and Parboil~\cite{Stratton2012ParboilAR}) and popular deep learning (DL) frameworks (i.e., 
TVM~\cite{ApacheTVM}, PyTorch~\cite{PyTorch}, TensorRT~\cite{TensorRT}, and FasterTransformer~\cite{FasterTransformer}).
Evaluation results demonstrate that {\sys} successfully identifies 11,745 idempotent instances
out of a total of 18,217 instances, with no false positives and a low false-negative rate of 18.54\%.
Impressively, all these GPU instances can be validated within 5\,{\us}, 
impacting the performance of applications by less than 1\%.
Furthermore, by utilizing the correct and efficient idempotency information provided by {\sys}, 
a GPU fault-tolerant system~\cite{Leng2020AsymmetricRE} can reduce the checkpoint cost of error-free execution from more than 115\% to less than 4\%, and 
a GPU scheduling system~\cite{Park2015ChimeraCP} can reduce the preemption latency by 84.2\%.

\vspace{0.5mm}
\stitle{Contributions.} We summarize our contributions as follows.
\begin{itemize}[leftmargin=*]
	\vspace{-2mm}
    \item The discovery of the existence and prevalence of {\pidemshort} GPU kernels in real-world applications.
    \item An in-depth study on idempotency that reveals the necessity and potential
    of instance-level idempotency validation.
    \item An approach to validate the instance-level idempotency at launching time 
	by checking memory access overlapping, along with optimizations enabling {\us}-scale validation.
    \item A prototype implementation and an evaluation that
    demonstrates the efficacy and efficiency of {\sys}.
\end{itemize}

\section{Background and Motivation} 
\label{sec:bg}

\subsection{Idempotence-based GPU Systems}
\label{subsec:bg-idem-sys}
Idempotent GPU kernels consistently produce the same output no matter where their execution 
is interrupted and then restart. This has been widely used to accelerate GPU applications 
where kernel execution may face either unintentional or intentional interruptions, 
as illustrated in the following cases.

\stitle{Case 1: Fault recovery.}
Tolerating transient faults, particularly bit flips~\cite{Wan2024MulBERRY,Li2024YesOM}, is crucial 
in safety-critical GPU applications~\cite{PrezCerrolaza2022GPUDF,Pattabiraman2020ErrorRM},
such as autonomous driving~\cite{Alcaide2018SafetyRelatedCA}.
A common approach to tolerate bit flips is to checkpoint the GPU memory periodically
and replay the program from the latest checkpoint upon fault~\cite{Huang2024PARALLELGPUOSAC,Parasyris2020CheckpointRS,Pea2015VOCLFTIT}.
However, this approach incurs significant performance overhead due to the extensive amount
of data that must be saved.
Therefore, prior work~\cite{Leng2020AsymmetricRE} leverages the idempotency of GPU kernels to 
reduce recovery overhead by only checkpointing GPU memory for non-idempotent kernels.
As expected, this optimization can work very well on 
applications where GPU kernels are mostly idempotent.

\stitle{Case 2: Preemptive scheduling.}
Latency-sensitive GPU applications, 
such as packet processing~\cite{Go2017APUNetRG,Kalia2015RaisingTB} and
real-time model inference~\cite{Howard2017MobileNetsEC,Fingler2023TowardsAM}, 
demand a fast kernel preemption mechanism for scheduling 
to meet the service level objective (SLO) of latency. 
However, the conventional preemption mechanism---context switching---introduces 
high preemption latency on the GPU (typically tens of microseconds) 
due to the large size of on-chip context (e.g., 20\,MB for NVIDIA V100~\cite{NVIDIAV100}). 
This may severely harm the responsiveness and safety of 
latency-critical GPU systems~\cite{Park2015ChimeraCP,Lin2016EnablingEP}.
Fortunately, researchers have recently 
leveraged the idempotency of GPU kernels to drastically reduce preemption latency~\cite{Park2015ChimeraCP,han2022reef,Lee2021IdempotenceBasedPG},
since idempotent kernels can be instantly killed (in about 1\,{\us}) 
and then restarted without saving their context.

\vspace{1.ex}
An important prerequisite for these idempotence-based systems to work efficiently is that
they need to accurately know the idempotency of GPU kernels.
We summarize three key requirements for the provided idempotency information.

\stitle{R0: Practicality (knowing before execution).}
The idempotency of a GPU kernel instance should be identified before actually executing it. 
This will allow idempotence-based systems to decide how to act on it based on its idempotency.

\stitle{R1: Correctness (no false positives).} 
A GPU kernel instance that is non-idempotent should not be mistakenly classified as idempotent. 
Otherwise, idempotence-based systems may re-execute a non-idempotent instance
from an inconsistent state, leading to incorrect results.

\stitle{R2: Efficiency (fewer false negatives).}
A GPU kernel instance that is idempotent should not be mistakenly classified as non-idempotent.
Otherwise, idempotence-based systems may miss the opportunity to leverage idempotence.


\subsection{Kernel-level Idempotency}

Prior work statically identifies a GPU kernel as either idempotent or
non-idempotent~\cite{Lee2021IdempotenceBasedPG,Leng2020AsymmetricRE}, and
treats all instances of a GPU kernel as idempotent or non-idempotent, respectively.
However, we discovered that a GPU kernel can have both idempotent instances and
non-idempotent instances, depending on the input of the instance.

\stitle{Motivating example.} 
Fig.~\ref{fig:vectoradd} shows a simplified version of the \code{vectorAdd} GPU kernel from 
CUDA samples~\cite{CUDASamples} that calculates the sum of two vectors.
When the kernel is invoked with three distinct pointers (i.e., the first instance with
arguments \code{x}, \code{y} and \code{z}), the GPU kernel instance is idempotent because 
the input buffers (i.e., \code{y} and \code{z}) are not overwritten during execution.
However, the second instance where the parameters \code{A} and \code{C} have the same value (i.e., \code{x})
is non-idempotent, because the input vector pointed by \code{x} is modified during execution, and 
the result of repeatedly executing this instance is different from that of only executing it once.

    
\stitle{Definition: kernel-level idempotency.} 
Considering that different instances of the same GPU kernel may vary in their idempotency 
(i.e., idempotent or non-idempotent), the idempotency of a GPU kernel can no longer be categorized 
as a simple binary distinction. 
In this paper, we propose a new categorization of \textbf{kernel-level idempotency} 
for GPU kernels based on the idempotency of their instances, as follows:
\begin{itemize}[leftmargin=*,leftmargin=15pt,itemindent=0pt]
    \vspace{-3pt}
    \item[{\fullcirc[0.7ex]}] \textbf{Idempotent} GPU kernel:
	All instances of the kernel are idempotent.
    \vspace{-3pt}
    \item[{\emptycirc[0.7ex]}] \textbf{Non-idempotent} GPU kernel:
	All instances of the kernel are not idempotent.
	\vspace{-3pt}
    \item[{\halfcirc[0.7ex]}] \textbf{{\pidemcap}} GPU kernel:
	Some instances of the kernel are idempotent, while others are not.
	\vspace{-3pt}
\end{itemize}

\noindent For example, as shown in Fig.~\ref{fig:vectoradd}, 
the \code{vectorSet} kernel 
is idempotent because the value of \code{A[idx]} remains the same 
no matter how many times it is re-executed.
In contrast, the \code{vectorInc} kernel 
is non-idempotent because the value of \code{A[idx]} is incremented by one
every time it is re-executed.
Lastly, the \code{vectorAdd} kernel 
is {\pidem}, as it can have both idempotent and non-idempotent instances.

\section{A Study of Idempotency in Applications}
\label{sec:study}
\label{subsec:bg-study}

\def\opensourcefootnote{\footnote{Though the framework of PyTorch and FasterTransformer is open-sourced, it
uses some closed-sourced GPU kernels (e.g., cuDNN~\cite{Chetlur2014cuDNNEP}).}}

\def\partialtracefootnote{\footnote{We also conducted the experiments on a 
RTX 3080 GPU (Ampere Architecture) and found 9 additional {\pidemshort} GPU kernels
that have both instances in the trace (i.e., for column \textup{\halfcirc[0.8ex]$_{\text{\textbf{/T}}}$}).}}

\def\trace3080footnote{\footnote{We also conducted these experiments on an NVIDIA RTX 3080 GPU, 
obtaining similar results that further confirmed our findings.}} 

To gain a better understanding of idempotency in real-world GPU applications, 
we conducted a comprehensive study on well-known GPU benchmarks 
(i.e., Rodinia~\cite{Che2009RodiniaAB} and Parboil~\cite{Stratton2012ParboilAR}) 
and popular DL frameworks, including TVM~\cite{ApacheTVM}, PyTorch~\cite{PyTorch}, 
TensorRT~\cite{TensorRT}, and FasterTransformer~\cite{FasterTransformer} (abbreviated as FT).
These experiments were performed on an NVIDIA GV100 GPU.\trace3080footnote 
More details can be found in $\S$\ref{sec:eval:setup}.

\stitle{Instance-level idempotency tracing.}
We build an idempotency tracing tool using GPU dynamic instrumentation~\cite{Villa2019NVBitAD}.
The tool records all accessed addresses on the GPU memory during execution.
We analyze the addresses to detect the presence of
a clobber anti-dependency access pattern, which specifically refers to a write-after-read without
a prior read-after-write on the same address~\cite{Kruijf2011IdempotentPA}.
If no clobber anti-dependency is identified, the instance is categorized as idempotent,
otherwise non-idempotent.

\stitle{Kernel-level idempotency labeling.}
It is non-trivial to correctly and efficiently identify idempotent or non-idempotent kernels, 
as it requires knowing the idempotency of all its instances.
However, it is straightforward to identify a {\pidemshort} kernel 
once both an idempotent and a non-idempotent instance are observed.
Therefore, we adopt a simple yet conservative approach to identify {\pidemshort} kernels.
We first try our best to find both idempotent and non-idempotent instances for a kernel 
to confidently label it as {\pidemshort}.
If this approach fails, we roughly label it as either idempotent or non-idempotent, 
although it is still possible that the kernel is actually {\pidemshort}.

Specifically, we first run the applications with public traces (i.e., traced instances)
and use our tracing tool to obtain the instance-level idempotency.
Then, we construct additional instances (i.e., generated instances)
that have opposite idempotency to their instances from the trace.
The construction is performed by changing the value of the pointers in the launch arguments.
If a traced instance is idempotent, we try to generate a non-idempotent instance by 
assigning the same value to the write pointers in the launch arguments.
Conversely, if a traced instance is non-idempotent, we try to generate an idempotent instance 
by assigning distinct values to all pointers in the launch arguments.
We use both traced and generated instances to label the kernel-level idempotency.

The main results for the idempotency of GPU kernels and instances
are reported in Table~\ref{tbl:kernels} and Table~\ref{tbl:instances}, respectively.
We outline three important observations as follows.

\begin{table}[t]
    \vspace{2mm}
    \centering
    \begin{minipage}{1.\linewidth}	
    \caption{\textit{\small{The kernel-level idempotency of the evaluated GPU kernels.
    The column \fullcirc[0.8ex], \emptycirc[0.8ex], and \halfcirc[0.8ex] indicate that 
	the GPU kernel is idempotent, non-idempotent, and {\pidemshort}, respectively.
    The column \textup{\halfcirc[0.8ex]$_{\text{\textbf{/{T}}}}$} indicates that 
	the GPU kernel has both idempotent and non-idempotent instances from their public application traces.
    }}}
    \label{tbl:kernels}
    \end{minipage} \\[2pt]
    \begin{minipage}{1.\linewidth}
    \ra{1.05}
    \centering
    \small{
    \begin{tabular}{@{}r|c|r|r r r r@{}}
    \toprule
    \textbf{GPU Apps}
    & \textbf{Code}
    & \textbf{\#Kernels} 
    & \fullcirc[0.8ex] & \emptycirc[0.8ex] & \halfcirc[0.8ex] & {~~}\halfcirc[0.8ex]$_{\text{\textbf{/{T}}}}$ \\
    \midrule
    Rodinia~\cite{Che2009RodiniaAB}       & {Source}  & {40}  & {7}  & {12} & {21}  & {2~}  \\
    Parboil~\cite{Stratton2012ParboilAR}  & {Source}  & {25}  & {4}  & {12} & {9}   & {0~}  \\
    TVM~\cite{ApacheTVM}                  & {Source}  & {308} & {0}  & {0}  & {308} & {0~}  \\
    PyTorch~\cite{PyTorch}                & {Binary}  & {66}  & {3}  & {1}  & {62}  & {2~}  \\
    TensorRT~\cite{TensorRT}              & {Binary}  & {58}  & {0}  & {2}  & {56}  & {0~}  \\
    FT~\cite{FasterTransformer} &{Binary} & {50} & {9} & {7} & {34}  & {0~}  \\
    \midrule 
    \multicolumn{2}{c|}{{All}}                        & {547} & {~}{23} & {~~}{34} & {490} & {4~} \\
    \bottomrule
    \end{tabular}
    }
    \end{minipage} \\[-15pt]
\end{table}

\def\codefootnote{\footnote{This GPU kernel is dynamically generated by PyTorch JIT module,
whose base template is \href{https:\/\/github.com\/pytorch\/pytorch\/blob\/664058fa83f1d8eede5d66418abff6e20bd76ca8\/aten\/src\/ATen\/native\/cuda\/CUDALoops.cuh\#L132}{here}.}}

\def\tracenote{\footnote{The detailed information of these four GPU kernels can be found
in Appendix A (see supplementary material).}}

\stitle{O1: The existence of {\pidemshort} kernels.}
The column \halfcirc[0.8ex]$_{\text{\textbf{/T}}}$ in Table~\ref{tbl:kernels} 
displays the number of kernels that include both idempotent and non-idempotent
instances from their public application traces. Upon analysis, we discovered 
four kernels with such instances, 
providing conclusive evidence of the existence of {\pidemshort} kernels.
For example, one of these kernels closely resembles the straightforward
\code{vectorAdd} kernel depicted in Fig.~\ref{fig:vectoradd}, performing an element-wise
tensor addition. These real-world cases highlight the risk of misclassifying non-idempotent
instances as idempotent (i.e., false positives) when treating {\pidem} kernels optimistically.
Consequently, the correctness of idempotence-based systems can be compromised,
directly contravening \textbf{R1}.

\stitle{O2: The majority of kernels are {\pidemshort}.}
Analyzing the column \halfcirc[0.8ex] in Table~\ref{tbl:kernels} reveals a total of
490 {\pidemshort} kernels, accounting for 89.6\% of all kernels. 
Among them, 486 kernels are labeled as {\pidemshort} because we constructed instances
that have the opposite idempotency to their instances from the trace. 
This indicates that simply modifying the input pointer of an instance can
alter its idempotency, further highlighting the prevalence of {\pidemshort} kernels.
These findings illustrate the necessity of cautiously treating {\pidemshort} GPU kernels 
as non-idempotent.
Although the correctness of idempotence-based systems would remain intact, 
most GPU kernels and instances would be erroneously classified as non-idempotent (i.e., false negatives), 
thereby compromising the efficiency of idempotence-based systems, contradicting \textbf{R2}.

Collectively, the observations presented above indicate that the existing 
approaches~\cite{Lee2021IdempotenceBasedPG,Leng2020AsymmetricRE,Park2015ChimeraCP} that
statically categorize kernels as either idempotent or non-idempotent and 
assume that idempotency for all instances is the same as the kernel-level idempotency, 
can greatly compromise the correctness (\textbf{R1}) or the efficiency (\textbf{R2}) 
of idempotence-based systems.

\begin{table}[t]
    \vspace{2mm}
    \centering
    \begin{minipage}{1.\linewidth}
    \caption{\textit{\small{The instance-level idempotency of the traced instances.
    The kernel-level idempotency is the same as that in Table~\ref{tbl:kernels}.
    The column \fullcirc[0.8ex]~{\normalsize{/\,\ding{72}}} (\emptycirc[0.8ex]~{\normalsize{/\,\ding{73}}})
    indicates the idempotent (non-idempotent) instances of idempotent (non-idempotent) kernels.
    The column \halfcirc[0.8ex]~{\normalsize{/\,\ding{72}}} and \halfcirc[0.8ex]~{\normalsize{/\,\ding{73}}}
    indicate the idempotent and non-idempotent instances of {\pidemshort} kernels, respectively.
    }}}
    \label{tbl:instances}
    \end{minipage} \\[2pt]
    \begin{minipage}{1.\linewidth}
    \ra{1.05}
    \centering
    \small{
    \begin{tabular}{@{}r|r|r|r|r@{~~~~}r@{~}}
    \toprule
    \textbf{GPU Apps}
    & \textbf{\#Instances} 
    & \fullcirc[0.8ex]~{\normalsize{/\,\ding{72}}} 
    & \emptycirc[0.8ex]~{\normalsize{/\,\ding{73}}} 
    & \halfcirc[0.8ex]~{\normalsize{/\,\ding{72}}} 
    & \halfcirc[0.8ex]~{\normalsize{/\,\ding{73}}} \\
    \midrule
    Rodinia~\cite{Che2009RodiniaAB}      & 4,527  & 85   &  78   & 4,334  & 30   \\ 
    Parboil~\cite{Stratton2012ParboilAR} & 1,033  & 103  &  738  & 192    & 0    \\ 
    TVM~\cite{ApacheTVM}                 & 609    & 0    &  0    & 609    & 0    \\ 
    PyTorch~\cite{PyTorch}               & 1,570  & 151   &  1   & 1,131  & 287  \\ 
    TensorRT~\cite{TensorRT}             & 478    & 0   &  8     & 465    & 5    \\ 
    FT~\cite{FasterTransformer}          & 10,000 & 229 & 988    & 7,119  & 1,664 \\
    \midrule
    \multicolumn{1}{c|}{{All}}           & 18,217  & 568  &  1,813  & 13,850  & 1,986  \\ 
    \bottomrule
    \end{tabular}
    }
    \end{minipage} \\[-15pt]
\end{table}

\stitle{O3: The majority of instances are idempotent.}
Table~\ref{tbl:instances} illustrates that out of all traced instances,
14,418 (79.1\%) are classified as idempotent (\ding{72}). 
Remarkably, even for {\pidemshort} kernels, 
13,850 out of 15,836 instances are idempotent (\halfcirc[0.8ex]\,{\normalsize{/\ding{72}}}), 
accounting for 87.5\% of their total instances. 
Consequently, it is evident that although {\pidemshort} kernels can have both
idempotent and non-idempotent instances, the majority of instances are indeed idempotent.
This indicates that determining the idempotency of each instance individually is 
a promising approach to satisfy the efficiency (\textbf{R2}) of idempotence-based systems
without compromising the correctness (\textbf{R1}).
However, accurately determining the idempotency of instances requires actually executing 
them, e.g., utilizing our tracing tool.
This contradicts \textbf{R0}.
 
\stitle{Our Insight.}
We found that the idempotency of the vast majority instances depends
solely on their launch arguments.
Specifically, in our study, simply by changing the value of the input pointers,
we constructed instances for {486} kernels with opposite
idempotency to their traced instances.
This presents a valuable opportunity to 
validate the instances' idempotency 
after they are launched at runtime but before actual execution (\textbf{R0}),
solely by knowing their launch arguments.

As a specific example, consider the \code{vectorAdd} 
GPU kernel in Fig.~\ref{fig:vectoradd}, once we know the addresses of the three buffers 
(\code{A}, \code{B} and \code{C}) and the parallelism (\code{bdim} and \code{gdim}),
we can determine the idempotency of the instance by validating whether the input buffers 
(\code{B} and \code{C}) might be overwritten.

\begin{figure}[t]
    \vspace{1mm}
    \begin{minipage}{1\linewidth}
    \centering\includegraphics[width=1.\linewidth]{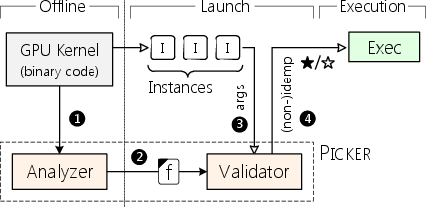}
    \end{minipage} \\[5pt]
    \begin{minipage}{1\linewidth}
    \caption{\emph{\small{Architecture of {\sys} and the workflow of using {\sys} 
    to correctly and efficiently run GPU kernels on idemp-based systems.}}}
    \label{fig:arch}
    \end{minipage} \\[-15pt]
\end{figure}

\section{{\sys}} 
\label{sec:design}

\def\extfootnote{\footnote{We discuss extending {\sys} to support multi-kernel idempotency 
and non-GPU-memory state in supplementary material.
}}

We present {\sys}, the first system that dynamically validates the
idempotency of GPU kernel instances at runtime.

\label{sec:overview}

Fig.~\ref{fig:arch} illustrates the architecture of {\sys} and its operation in
conjunction with an idempotence-based system.
{\sys} comprises two components: a static analyzer and a runtime validator.
The analyzer is responsible for analyzing the binary code of GPU kernels offline (\ding{202})
to generate functions for validating the idempotency of GPU kernel instances.
These functions are then compiled as dynamic libraries and are loaded by the
validator during runtime (\ding{203}).
Each time the application launches an instance, {\sys} intercepts the instance and
validates its idempotency using the launch arguments (\ding{204}).
The instance is then labeled with its instance-level idempotency and
passed to the GPU runtime for execution. 
The idempotence-based system can then take appropriate action for the 
instance based on the validation result (\ding{205}).

\subsection{Validate Idempotency by Checking R/W Overlaps}
\label{subsec:what}

To fulfill the requirements of idempotence-based systems,
we should first establish a condition that can be validated prior to
the execution of a GPU kernel instance (\textbf{R0}), and is sufficient
for an instance to be considered idempotent.

An idempotent instance ensures that, when re-executed with the same inputs 
as its initial execution, the output will be consistent with the exactly-once execution.
Therefore, a common approach is to check for modifications to the \emph{input memory}
of an instance during execution. 
Input memory refers to the memory region that was first accessed as a read.
Specifically, an instance is considered as non-idempotent 
if there exists any write-after-read access patterns
on its input memory~\cite{Lee2021IdempotenceBasedPG,Kruijf2012StaticAA}.
However, analyzing the memory access order in a GPU kernel instance, which utilizes 
multiple concurrent GPU threads and has interleaved memory accesses, is extremely complex, 
if not impossible.

\stitle{Our approach: validating R/W overlaps}.
{\sys} enforces a more stringent condition to validate the idempotency of a
GPU kernel instance, where an instance is considered non-idempotent if there exists any
overlap in the read and write addresses of all potential memory accesses, regardless of the
access order. On the other hand, an instance is classified as idempotent if
{\sys} can ensure that each byte of GPU memory is either read-only or write-only.
Although this may occasionally lead to false negatives (e.g., a read-after-write
pattern misclassified as non-idempotent), 
it upholds the requirement of no false positives (\textbf{R1}).

Up to this point, the problem of validating instance-level idempotency has transformed 
into the task of accurately predicting the memory addresses that 
the instance will access.
To satisfy \textbf{R1}, the predicted addresses must cover all the really 
accessed addresses.
To satisfy \textbf{R2}, the predicted addresses should minimize the
inclusion of non-accessed addresses.

\subsection{Strawman}\label{subsec:strawman}

We start by presenting a strawman solution, to demonstrate the fundamental design of {\sys}.
The basic idea is to \emph{simulate} the execution of the GPU kernel instance
by using the launch arguments after the instance is launched by the application, allowing us
to obtain the concrete addresses of all accesses on the GPU memory.
Specifically, {\sys} divides the idempotency determination into two phases: 
\emph{static analysis} and \emph{runtime validation}.
The static analysis is performed offline to generate functions that are utilized to
calculate memory addresses.
These functions are then invoked at runtime using the launch arguments of the instance,
to calculate the addresses and validate the idempotency of the instance.

\stitle{Static analyzer.}
The analyzer in {\sys} first analyzes the binary code of GPU kernels via symbolic execution.
The symbolic execution engine records the memory addresses when it encounters load/store 
instructions on GPU global memory.
Each recorded address represents a symbolic expression, referred to as a \emph{symbolic address}.
The variables inside a symbolic address typically consist of the parameters of the GPU 
kernel (including the parallelism parameters), as well as the block and thread IDs of the
GPU threads that execute the instruction.
Moreover, each symbolic address is associated with a list of path conditions that must be
satisfied to traverse the path during symbolic execution.
Finally, the analyzer saves the symbolic addresses, which will be loaded by the validator 
at runtime. Additionally, the analyzer also generates a \emph{global condition} that contains
some preconditions (explained later).

\begin{figure}[t]
\vspace{1mm}
\begin{minipage}{1.\linewidth}
\centering\includegraphics[width=1.\linewidth]{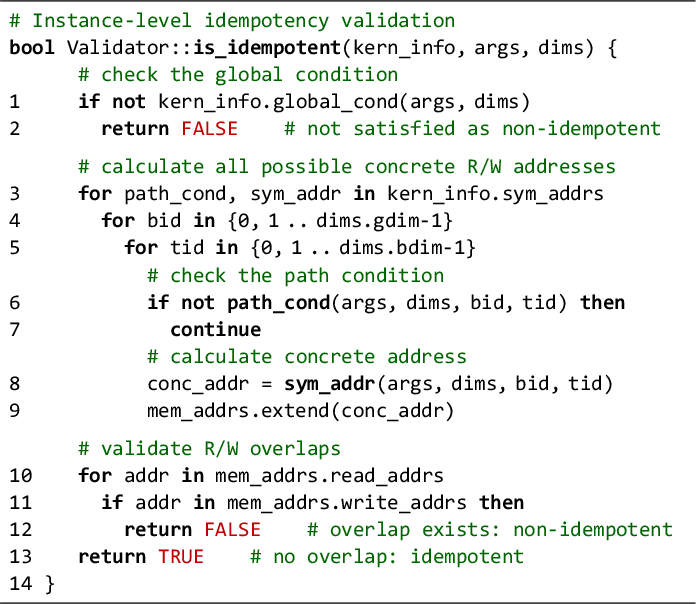}
\end{minipage} \\[10pt]
\begin{minipage}{1.\linewidth}
\caption{\emph{\small{Pseudocode of dynamic idempotency validation.}}}
\label{fig:strawman}
\end{minipage}  \\[-25pt]
\end{figure}

\stitle{Runtime validator.}
The validator in {\sys} is responsible for validating the idempotency of instances at runtime,
using the symbolic addresses generated by the analyzer.
Fig.~\ref{fig:strawman} shows the simplified pseudocode of the runtime idempotency validation.
For each GPU kernel instance, the validator first validates the global condition (Lines 1--2),
and then calculates the concrete addresses of all memory accesses (Lines 3--9).
For each symbolic address, the validator iterates through all block and thread IDs to calculate the 
concrete address accessed by each GPU thread (Lines 4--5).
For simplicity but without loss of generality, we only consider that the block and thread IDs
are one-dimensional.
Within each iteration, the validator confirms whether the path condition of the
symbolic address is satisfied (Lines 6--7), and then calculates the concrete address (Line 8).
Finally, the validator determines the idempotency of the instance by checking the overlap
between the read and write addresses (Lines 10--12).

In order to satisfy \textbf{R1}, {\sys} must analyze all possible memory addresses
an instance may access. This requires the analysis to cover all possible control
flows and data flows. Because GPU kernels typically have no complex 
behaviors~\cite{Mai2023HONEYCOMB,Kamath2021iGUARDIA}, the analysis can be 
gracefully completed most of the time. 
However, there are still some cases where the analysis may fail to cover all flows,
e.g., due to indirect function calls.
For these cases, {\sys} conservatively assumes that the kernel is non-idempotent 
for all its instances.

\etitle{Handling unbounded loops.}
The symbolic execution may also fail due to unbounded loops.
When encountering an unbounded loop, the analyzer executes the loop body a fixed number
of times, then breaks the loop. 
If the loop is actually executed more times than the analyzer expected, the instance
should be treated as non-idempotent.
The analyzer achieves this by generating a global condition that the instance 
must satisfy if the loop is executed more times than expected.
The validator then checks whether the condition is satisfied. 
If not satisfied, the instance is classified as non-idempotent.

\etitle{Handling non-parameter variables.}
A symbolic address may consist of variables that are not from the parameters,
dimensions or IDs, called \textit{non-parameter variables}.
For example, dereferencing a double pointer (e.g., $\code{A[0][0]}$) may generate a symbolic address that
consists of a variable read from the GPU memory.
It is worth noting that the validator can only get the concrete values of the parameters
and dimensions of the instance, but not the variables from the GPU memory.
Therefore, {\sys} conservatively assumes that these variables
could be any concrete value. Most of the time, a symbolic address with non-parameter variables is 
considered to be overlapped with any other address, and thus the instance is classified 
as non-idempotent.

\etitle{Idempotent/non-idempotent GPU kernels.}
{\sys} makes a simple analysis of the symbolic addresses to detect kernels whose instances
are always idempotent or non-idempotent. For example, if a kernel has no load or store instructions,
it can be safely treated as idempotent. While, if a kernel reads and
writes the same symbolic address, or contains an atomic instruction, it is also directly classified
as non-idempotent. The validator directly returns kernel-level idempotency without performing the
validation for these kernels.

\begin{figure}[t]
\vspace{1mm}
\begin{minipage}{1.\linewidth}
\centering\includegraphics[width=1.\linewidth]{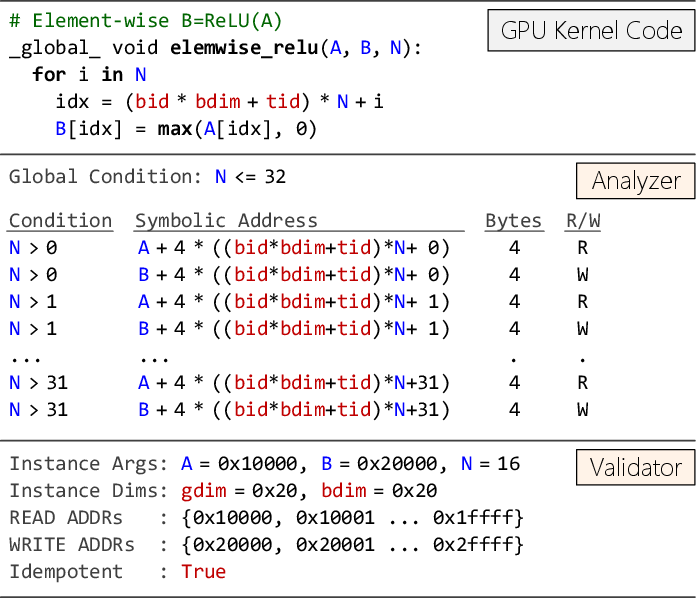}
\end{minipage} \\[10pt]
\begin{minipage}{1.\linewidth}
\caption{\emph{\small{An example for idempotency validation using our strawman solution. 
  \code{elemwise\_relu} is a simplified version of a {\pidemshort} GPU kernel
  mentioned in $\S$\ref{subsec:bg-study} from PyTorch~\cite{PyTorch}.}}}
\label{fig:strawman-example}
\end{minipage}  \\[-15pt]
\end{figure}

\stitle{Example.} 
Fig.~\ref{fig:strawman-example} illustrates a kernel performing the
ReLU activation function on each element of the input array, with each GPU thread responsible
for computing \code{N} elements. 
The analyzer symbolically executes the code and records the addresses
of all memory accesses. Each symbolic address consists of the parameter 
variables (i.e., \code{A}, \code{B}, and \code{N}), the dimension variables (i.e., \code{bdim} and \code{gdim}),
alongside the block and thread IDs (i.e., \code{bid} and \code{tid}).
Because the kernel includes an unbounded loop, the analyzer
only executes the loop body for a predefined number of times (i.e., 32 times in the example), 
and generates a global condition (\code{N<=32}) to identify instances that execute the loop
more times.
At runtime, the validator first validates the global condition.
It then enumerates all \code{bid} and \code{tid} to calculate the addresses.
Ultimately, the validator identifies no overlap in the read and write addresses,
classifying the instance as idempotent.

\section{Achieving {\us}-scale Validation Latency}\label{sec:opt}

Unlike the kernel-level approach, where all classifications are
done offline, {\sys} introduces runtime overhead due to dynamic validation
in the critical path of GPU kernel launches.
It typically takes a few microseconds 
to start execution after being launched. 
Therefore, {\sys} must complete the validation
in microseconds to avoid non-negligible overhead.

However, our experimental results (see $\S$\ref{sec:eval:opt}) show that the strawman solution
takes up to 50\,ms to validate a GPU kernel instance, which is four orders of magnitude higher
than the required latency in the microsecond scale, 
introducing a significant overhead for the application.

Upon examining the pseudocode in Fig.~\ref{fig:strawman}, we can estimate the
validation latency as the product of the number of symbolic addresses
in the kernel ($N_{symAddr}$), the number of GPU threads for the
instance ($N_{Thrd}$), and the average execution time to calculate the concrete
value of a symbolic address ($T_{Instr}$). 
Each of these three factors significantly influences the validation latency.
Therefore, we propose several optimizations to reduce these factors
individually, with the aim of achieving the microsecond-scale validation latency.

\subsection{Reducing $N_{Thrd}$: Range-based Address Model}
\label{subsec:opt-range}

Our strawman solution performs concrete address calculations for
each GPU thread, resulting in the execution time being proportional to the
total number of GPU threads ($\code{bdim}*\code{gdim}$).
However, an instance typically comprises thousands of GPU threads,
leading to lengthy computation times.
Therefore, it is necessary to reduce this factor to a constant value that
is independent of the number of GPU threads.

\stitle{Representing addresses with ranges.}
We observed that the memory addresses accessed by different GPU threads
at the same load/store instruction are often \textit{consecutive}.
This pattern is recommended to enhance memory access efficiency by coalescing
the memory accesses of GPU threads~\cite{Che2011DymaxionOM,Jang2011ExploitingMA}. 
Consequently, instead of enumerating all the individual addresses,
{\sys} represents the memory addresses accessed by all GPU threads at the same
load/store instruction using a \textit{range}, which is defined by 
two values: the upper bound and the lower bound.
If we can calculate the two bounds of the range in constant time,
the execution time of the validation would be independent of the factor $N_{Thrd}$.
However, without enumerating all the GPU thread IDs,
computing the lower bound, for instance, requires solving the analytic form of the bound function in advance.
This function, denoted as 
$LB(args) = \mathop{\min}_{bid,tid} Addr(bid, tid, args)$, 
represents a non-trivial challenge as it involves solving bound functions for
symbolic addresses of arbitrary form.


\stitle{Estimating range bounds with monotonicity.}
We further discovered that in most cases (82\% symbolic addresses of the evaluated
kernels), the values of symbolic addresses increases
monotonically with respect to GPU thread IDs (\code{bid} and \code{tid}).
Taking the symbolic address $\code{A+4*(bid*bdim+tid)*N}$ in Fig.~\ref{fig:strawman-example}
as an example, with concrete values, the 
symbolic address becomes $\code{0x10000+4*(bid*0x20+tid)*16}$,
which demonstrates a monotonically increasing trend with respect to 
both \code{bid} and \code{tid}.
In other words, the lowest address is accessed by the thread with the smallest IDs (\code{bid} and \code{tid}), 
and the highest by the largest.
This monotonicity allows us to calculate the address range using only the smallest 
and largest thread IDs, making validation latency independent of $N_{Thrd}$.

\stitle{Checking monotonicity.}
However, the monotonicity of symbolic addresses is not always guaranteed.
Before using the smallest and largest GPU thread IDs to calculate the range, 
{\sys} statically checks the monotonicity of the 
symbolic addresses on the variables of GPU thread IDs at offline. 
To check the monotonicity of $Addr(bid, tid, args)$ on $bid$, {\sys} exploits
the capabilities of the SMT solver~\cite{Moura2008Z3AE} to prove that 
$Addr(x, tid, args) \ge Addr(y, tid, args)$ holds true whenever $x > y$.
If the solver cannot find a counterexample after exhausting the entire search space,
we say that the symbolic address is monotonically increasing with respect to $bid$.
In contrast, if the solver finds a counterexample or times out, the symbolic address
is considered to be non-monotonic. 
The checking is also performed on $tid$ in the same way.

In fact, the monotonicity is often satisfied conditionally. 
For instance, in the first symbolic address in Fig.~\ref{fig:strawman-example},
if \code{A} is a very large address (e.g., $2^{64}$$-1$), 
the address may overflow when \code{bid} and \code{tid} increase, making it not monotonic. 
But real-world applications rarely encounter such overflow,
because the address space is much smaller than $2^{64}$.
Therefore, {\sys} makes prior constraints on the value of the launch arguments
for the monotonicity checking. These constraints are also added
to the \textit{global condition} to be validated at runtime.
For example, {\sys} may assume $\code{A}<2^{56}$, $\code{N}<2^{5}$, $\code{bdim}<2^{10}$,
and $\code{gdim}<2^{10}$, which are reasonable constraints for real-world applications and 
makes symbolic addresses monotonic. If an instance violates these constraints, {\sys} would
conservatively treat it as non-idempotent.

The constraints can be either inferred by {\sys} or manually specified
by the programmer. Currently, {\sys} can automatically generates constraints for
pointer-typed parameters. 
Other types of parameters must be specified manually.

\stitle{Handling non-monotonic symbolic addresses.}
{\sys} attempts to transform a non-monotonic symbolic address into a monotonic form by
replacing certain subexpressions with new independent variables.
For example, the symbolic address $\code{A+tid\%10}$ is non-monotonic with respect
to \code{tid}, but it can be transformed into $\code{A+var}$
where \code{var} is a new variable constrained to be in the range $[0, 9]$. 
This guarantees the transformed address's range is a subset of the original, avoiding false
positives (\textbf{R1}), but may lead to false negatives.

The transformation involves two steps. First, {\sys} traverses the expression tree
of the symbolic address bottom-up to find non-monotonic subexpressions. 
Then, it uses SMT solver to determine the minimum and maximum values of these non-monotonic
subexpressions, and replaces them with new variables constrained to the range.
The new symbolic address is then checked for monotonicity, and the process is repeated
until the it is monotonic.
For most of the cases, the transformation can stop after eliminating periodic functions
like $\code{tid\%10}$ and $\code{tid\&7}$. In the worst case, the whole symbolic address
may be replaced with a new variable, which is almost equivalent to
classifying the kernel as non-idempotent.

\stitle{Leveraging constraints on GPU thread IDs.}
Using the smallest and largest GPU thread IDs to calculate the range of a monotonic
symbolic address may overestimate the actual accessed address range if the instruction
is not executed by all GPU threads. 
For instance, if a symbolic address has a path condition of $\code{tid<10}$, but is launched
with $\code{bdim=32}$, the address should only be calculated with \code{tid} in the 
range $[0, 9]$, not $[0, 31]$.
Therefore, {\sys} should leverage these path conditions to narrow down the estimated range.

{\sys} utilizes pattern matching to identify path conditions that can
narrow down the range. It first looks for conditions that take the form
of a comparison between a variable part (an expression with GPU thread IDs)
and a constant part (an expression consisting solely of launch arguments
and constants). It then replaces the variable part with
a new variable whose range is restricted by the constant part.
For example, the path condition $\code{tid<}\code{N}$ matches this pattern, where the
variable part is \code{tid} and the constant part is \code{N} which is a launch argument.
The range of the variable part can be confined to $[0, min(\code{N-1},\code{bdim-1})]$,
with the comparison between $\code{N-1}$ and $\code{bdim-1}$ left to the runtime validator.

\subsection{Reducing $N_{symAddr}$: Range Compaction}\label{subsec:opt3}

The presence of loops causes load/store instructions to be symbolically
executed multiple times, generating a large number of symbolic addresses
that increases with the number of loop iterations, leading to a significant value of $N_{symAddr}$.
The primary approach to reducing $N_{symAddr}$ is to merge symbolic addresses generated from the
same instruction within a loop during the analysis, resulting in a single merged symbolic address
to calculate at runtime.

We observed that addresses of the load/store instructions inside loops in most kernels
consist of only loop-invariant variables and induction variables. 
Induction variables are variables that are incremented or decremented by a fixed step in each
loop iteration~\cite{Haghighat1995}.
For example, all kernels generated by TVM~\cite{ApacheTVM} that we evaluated match this pattern.
Our strawman solution which unrolls the loop, is essentially equivalent to enumerating the values
of induction variables.
However, it is unnecessary to enumerate the induction variables, as their range can be
symbolically analyzed using the initial value, step value, and the loop iterations.
Thus, we can treat induction variables as another type of independent variable similar 
to GPU thread IDs that fit the range-based address model.

Specifically, {\sys} performs dataflow analysis to identify the induction variables and
loop-invariant expressions offline.
If a symbolic address consists of only induction variables and loop-invariant expressions,
and the range of induction variables can be represented using only launch arguments,
{\sys} unrolls the loop body once and retains the induction variables in
the symbolic addresses. These variables are then included in the monotonicity
checking and range calculation.

For example, consider the loop in Fig.~\ref{fig:strawman-example} with a read of
the address of $\code{A}[\code{(bid*bdim+tid)*N+i}]$. 
Here, \code{i} is as an induction variable incrementing by 1 in each loop iteration, while the rest variables are
loop-invariant.
{\sys} can statically determine that \code{i} iterates from 0 to \code{N-1}.
Our strawman solution unrolls the loop and generates 32 symbolic addresses for this instruction.
However, {\sys} retains \code{i} as an independent variable in the symbolic address.
It allows \code{i} to further participate in the monotonicity checking,
and the calculation of the concrete range using its bound values (i.e., 0 and \code{N-1}).
his optimization reduces the number of symbolic addresses generated for the instruction to only one, and for the 
entire kernel to only two, significantly reducing $N_{symAddr}$.

\def\IVfootnote{\footnote{An induction variable is a variable that is incremented 
or decremented by a fixed step in each iteration of the loop, or a linear function of 
another induction variable.}}

\subsection{Reducing $T_{Instr}$: Compiled Execution}\label{subsec:opt4}
A straightforward approach to calculate the concrete ranges of symbolic addresses
is to directly use the \textit{substitute} function of the symbolic engine.
This function replaces symbolic variables with concrete values and evaluates
the expressions. Nevertheless, the interpretive evaluation of symbolic expressions
is considerably slow. In contrast, {\sys} converts the symbolic addresses into 
native code, specifically C language, and subsequently compiles it into
a dynamic shared library.

{\sys} also applies several common optimizations to the native code execution.
For example, the loop over the symbolic addresses is unrolled, and the functions that 
calculate the ranges are all inlined. This enables the compiler to perform
powerful compilation optimizations, such as common expression extraction, to further
reduce validation latency.

\section{Discussion} 

\subsection{Extensions of {\sys}}
We discuss how to extend {\sys} to support other use cases.

\stitle{Multi-kernel idempotency.}
This paper only discusses the idempotency of a single GPU kernel instance.
But the key mechanism behind {\sys}---the ability of predicting read and write
addresses of a GPU kernel instance---can be extended to support multi-kernel idempotency.
For example, {\sys} can be extended to validate the idempotency of a list of sequentially executed 
GPU kernel instances with two steps. First, {\sys} predicts the read and write addresses
of each GPU kernel instance. Second, {\sys} sorts the instances by their launch order,
and then checks the clobber anti-dependency across the instances.
{\sys} can also be extended to validate the idempotency of a list of 
concurrently executed GPU kernel instances, by simply checking the overlap
of read and write addresses among all concurrent instances.

\stitle{Non-GPU-memory states.}
This paper assumes that the GPU kernel instance only reads and writes the 
GPU memory. Some GPU applications may also directly access other system states,
such as the CPU memory\cite{Jeong2023FastAE}, disk\cite{Silberstein2013GPUfsIA}, 
and persistent memory\cite{Pandey2022GPMLP}.
Thanks to the virtual memory system, all these system states are also mapped 
to the virtual address space, and thus can be handled by {\sys} in a similar way.
However, an important prerequisite is that one physical address can only be
mapped to at most one virtual address, otherwise {\sys} may generate false positives.


\subsection{False-negative Cases}

\begin{figure}[t]
    \vspace{1mm}
    \begin{minipage}{1.\linewidth}
    \centering\includegraphics[width=1.\linewidth]{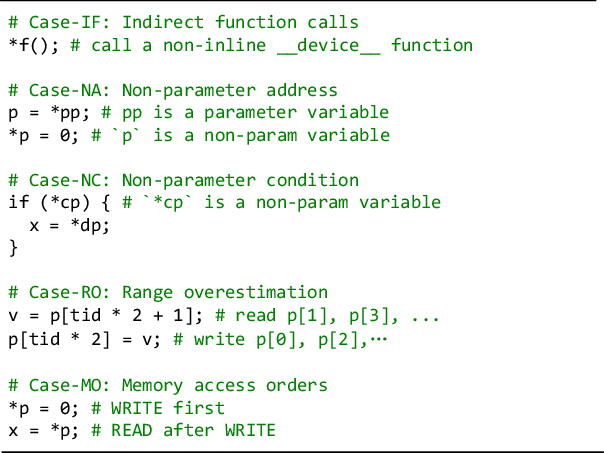}
    \end{minipage} \\[10pt]
    \begin{minipage}{1\linewidth}
    \caption{\textit{\small{Examples of false-negative cases.}}}
    \label{fig:false-example}
    \end{minipage} \\[-15pt]
\end{figure}

{\sys} is designed to meet the no-false-positive goal at first, but 
it may provide false-negative results in some cases.
In this section, we discuss the cases of false negatives and the possible solutions.
Fig.~\ref{fig:false-example} shows the examples of false-negative cases.

\stitle{Indirect function call (IF).}
{\sys} performs symbolic execution on GPU kernels at offline, which currently does
not support GPU kernels calling device functions indirectly (e.g., via function pointers).
A possible solution to support indirect function calls is to transform the indirect
function calls into a switch statement among all possible callees, and then perform
symbolic execution on each callee. We leave it as a future work.

\stitle{Non-parameter address (NA).}
Currently, {\sys} can only handle the memory access instructions whose addresses
only depend on the launch arguments, dimensions and thread IDs.
If n symbolic address consists of non-parameter variables, {\sys} conservatively
assumes that they could be any concrete value, which overlaps with any other ranges.
There are typically three types of non-parameter variables 
(1) a variable read from global memory (e.g., \code{*pp} in Fig.~\ref{fig:false-example}),
(2) a variable read from shared memory (i.e., variables defined with \code{__shared__}),
and (3) a variable read from non-parameter constant memory (i.e., variables defined with \code{__const__}).
A possible solution is to allow users to provide the possible values of these variables,
which we leave as a future work.

\stitle{Non-parameter condition (NC).}
The path constraint of a symbolic address may also contain non-parameter
variables (e.g., \code{if (*cp)} in Fig.~\ref{fig:false-example}).
{\sys} conservatively assumes that the non-parameter condition is always true 
to cover all possible memory access paths, which may lead to false negatives
if the condition is actually false for an instance.
Similar to the case of non-parameter addresses, a possible solution is to allow users
to provide the possible values of these variables, which we leave as future work.

\stitle{Range overestimation (RO).}
The range-based address model may amplify the set of real accessed addresses, 
and thus causes false overlapping.
The amplification can be caused by two reasons. 
First, the symbolic address may be not consecutive.
For example, as shown in Fig.~\ref{fig:false-example} the read set \{1, 3, 5\} is
not overlap with the write set \{0, 2, 4\}, but using the range-based model
creates one (i.e., the ranges [1, 5] and [0, 4] overlap).
A possible solution is to detect such patterns during static analysis, which we leave as
a future work.
Second, {\sys} may overestimate the bound of a symbolic address. 
For example, for a monotonic symbolic address, {\sys} may fail to analyze the 
constraints on \code{tid}, and conservatively uses \code{0} and \code{bdim-1}
as the lower and upper bound of \code{tid} respectively.
A possible solution is to conduct a more precise constraints analysis on \code{tid}
and \code{bid}, which we leave as future work.

\stitle{Memory-access orders (MO).}
{\sys} validates the idempotency of a GPU kernel instance by checking the overlap 
of read and write addresses, without considering the memory access orders.
A read-after-write pattern on the same address is actually idempotent, but {\sys}
will classify it as non-idempotent.
A possible solution is to statically analyze the memory access orders, with the 
consideration of the synchronization among GPU threads, which we leave as a future work.

\section{Evaluation}\label{sec:eval}

\begin{table*}[t]
\vspace{2mm}
\centering
\begin{minipage}{1\linewidth}
\caption{\emph{\small{The dynamic validation results of idempotency 
for GPU kernels and instances in GPU applications.
The column \fullcirc[0.8ex], \emptycirc[0.8ex], and \halfcirc[0.8ex] indicate that 
the GPU kernel is idempotent, non-idempotent, and {\pidem}, respectively.
The column {\normalsize{\ding{72}}} and {\normalsize{\ding{73}}} indicate that 
the GPU kernel instance is idempotent and non-idempotent, respectively.
\textup{\textbf{F+}} and \textup{\textbf{F--}} indicate false positives and false negatives, 
respectively. 
}}}
\label{tbl:acc}
\end{minipage}  \\[2pt]
\begin{minipage}{1\linewidth}
\ra{1.05}
\centering
\small{
\begin{tabular}{@{~}r @{~~~}r@{~~~}r rrr rrr rr rr@{~}r@{~}rr@{}}
\toprule
\multirow{2}{*}{\textbf{GPU Apps}}
& \multirow{2}{*}{\textbf{\#Kernels}} 
& \multirow{2}{*}{\textbf{\#Instances}} 
& \multicolumn{3}{c}{\textbf{Manual}} 
& \multicolumn{3}{c}{\textbf{Kernel-level}} 
& \multicolumn{2}{c}{\textbf{Truth}} 
& \multicolumn{5}{c}{\textbf{Instance-level}} \\
\cmidrule(lr){4-6} \cmidrule(lr){7-9} \cmidrule(lr){10-11} \cmidrule(lr){12-16}
& &  
& \fullcirc[0.85ex] & \emptycirc[0.85ex] & \halfcirc[0.85ex]
& \fullcirc[0.85ex] & \emptycirc[0.85ex] & \halfcirc[0.85ex]
& {\normalsize{\ding{72}}} & {\normalsize{\ding{73}}} 
& {\normalsize{\ding{72}}} & {\normalsize{\ding{73}}} & {\textbf{F--}} & {\textbf{(\%)}} & {\textbf{F+}~~} \\
\midrule
Rodinia~\cite{Che2009RodiniaAB} 
& 40   & 4,527      
& 7      & 12    & 21     
& 4      & 23    & 13     
& 4,419  & 108          
& 4,033  & 494   & 386   & (8.74)  & 0~~    \\
Parboil~\cite{Stratton2012ParboilAR} 
& 25   & 1,033	    
& 4      & 12    & 9      
& 1      & 15    & 9      
& 295    & 738          
& 222    & 811   & 73    & (24.75) & 0~~    \\
TVM~\cite{ApacheTVM} 
& 308  & 609	    
& 0      & 0     & 308    
& 0      & 0     & 308    
& 609    & 0            
& 596    & 13     & 13    & (2.13)  & 0~~    \\
PyTorch~\cite{PyTorch} 
& 66   & 1,570      
& 3      & 1     & 62     
& 3      & 22    & 41     
& 1,282  & 288
& 824    & 746   & 458   & (35.73) & 0~~    \\
TensorRT~\cite{TensorRT} 
& 58   & 478	    
& 0      & 2     & 56     
& 0      & 17    & 41     
& 465    & 13      
& 267    & 211   & 198   & (42.58) & 0~~    \\
FT~\cite{FasterTransformer}
& 50   & 10,000
& 9 & 7 & 34
& 8 & 19 & 23
& 7,348 & 2,652
& 5,803 & 4,197 & 1,545 & (21.03) & 0~~ \\
\midrule
\multicolumn{1}{c}{{All}} 
& 547  & 18,217	    
& 23      & ~~34  & 490    
& 16     & ~~96  & 435    
& 14,418  & 3,799   
& 11,745  & 6,472 & ~~2,673 & (18.54) & 0~~    \\
\bottomrule
\end{tabular}
}
\end{minipage} \\[-10pt]
\end{table*}

\begin{table}[t]
    \vspace{2mm}
    \centering
    \begin{minipage}{1\linewidth}
    \caption{\emph{\small{The breakdown of GPU kernels that {\sys} identifies as non-idempotent
    for various reasons.
    The column \emptycirc[0.8ex] indicates that the kernel is non-idempotent.
    \textup{\textbf{IF}}, \textup{\textbf{PE}}, \textup{\textbf{NA}}, and \textup{\textbf{SO}} 
    are abbreviations for ``indirect function call'', ``path explosion'', ``non-parameter address'',
    and ``symbolic overlap'', respectively.
    }}}
    \label{tbl:nonidem}
    \end{minipage} \\[2pt]
    \begin{minipage}{1.\linewidth}
    \ra{1.05}
    \centering
    \small{
    \begin{tabular}{@{~}r | r | r@{}r@{}r | r r r r@{~}}
    \toprule
    {\textbf{GPU Apps}~} & {\textbf{\#Kernels}}  & & \multicolumn{1}{c}{\emptycirc[0.85ex]} &
    & \textbf{~~~~IF} & \textbf{PE} & \textbf{NA}  & \textbf{SO} \\
    \midrule
    Rodinia~\cite{Che2009RodiniaAB}~       & 40   & & 23~~ &  & 7   & 2  & 3  & 11   \\
    Parboil~\cite{Stratton2012ParboilAR}~  & 25   & & 15~~ &  & 1   & 1  & 3  & 10   \\
    TVM~\cite{ApacheTVM}~                  & 308  & & 0~~  &  & 0   & 0  & 0  & 0    \\
    PyTorch~\cite{PyTorch}~                & 66   & & 22~~ &  & 0   & 16 & 5  & 1   \\
    TensorRT~\cite{TensorRT}~              & 58   & & 17~~ &  & 0   & 12 & 4  & 1   \\
    FT~\cite{FasterTransformer}~           & 50   & & 19~~ &  & 0   & 12 & 4  & 3   \\
    \midrule
    \multicolumn{1}{c|}{{All}}             & 547  & & 96~~ &  & 8  & 43  & 19  & 26  \\
    \bottomrule
    \end{tabular}
    } 
    \end{minipage} \\[-15pt]
\end{table}

\begin{table}[t]
    \vspace{2mm}
    \centering
    \begin{minipage}{1\linewidth}
    \caption{\emph{\small{The breakdown of GPU kernel instances that {\sys} incorrectly validates 
    as non-idempotent (false negatives) for various reasons.
    \textup{\textbf{F--}} indicates false negatives in the validation of GPU kernel instances.
    \textup{\textbf{IF}}, \textup{\textbf{PE}}, \textup{\textbf{NA}}, \textup{\textbf{NC}}, 
    and \textup{\textbf{RO}} are abbreviations for ``indirect function call'', 
    ``path explosion'', ``non-parameter address'', ``non-parameter condition'', and 
    ``range overestimation'', respectively.
    }}}
    \label{tbl:fnreason}
    \end{minipage} \\[2pt]
    \begin{minipage}{1.\linewidth}
    \ra{1.05}
    \centering
    \small{
    \begin{tabular}{@{~}r |r@{~}r@{~}r@{~}|@{~}r r r r r@{~}}
    \toprule
    {\textbf{GPU Apps}} & & {\textbf{F--}} &
    & \textbf{~~~~~IF} & \textbf{~~PE} & \textbf{~~NA} & \textbf{~~NC} & \textbf{~~RO}   \\
    \midrule
    Rodinia~\cite{Che2009RodiniaAB}       && 386 & & 44   & 15  & 71     & 1   & 255    \\
    Parboil~\cite{Stratton2012ParboilAR}  && 73  & & 0    & 12  & 59     & 0   & 2     \\
    TVM~\cite{ApacheTVM}                  && 13  & & 0    & 0   & 3      & 0   & 10     \\
    PyTorch~\cite{PyTorch}                && 458 & & 0    & 119 & 314    & 1   & 24    \\
    TensorRT~\cite{TensorRT}              && 198 & & 0    & 117  & 62      & 0   & 19    \\
    FT~\cite{FasterTransformer}           && 1,545 & & 4  & 263 & 1,214     & 64  & 0     \\
    \midrule
    \multicolumn{1}{c|}{{All}}             && 2,673 & & 48  & 526 & 1,723  & 66  & 310 \\
    \bottomrule
    \end{tabular}
    }
    \end{minipage} \\[-15pt]
\end{table}

We implemented {\sys} with 13\,K lines of Python for the static analyzer
and 1\,K lines of C++ for the runtime validator.
{\sys} is implemented to analyze the assembly code of GPU kernels rather than source code,
allowing it to handle closed-source kernels (e.g., from cuDNN~\cite{Chetlur2014cuDNNEP}) 
and dynamically generated ones (e.g., from PyTorch~\cite{PyTorch}). 
Currently, {\sys} supports the ISA of NVIDIA 
Volta and Ampere GPUs (e.g., GV100 and RTX 3080 in our experiments).
We omit similar results on RTX 3080 due to space limitation.
The symbolic execution engine is based on claripy~\cite{Shoshi2016SoK} and
Z3~\cite{Moura2008Z3AE}.

\subsection{Experimental Setup}
\label{sec:eval:setup}

\noidentstitle{Testbed.}
All experiments were conducted on a PC equipped with one Intel i7-13700 CPU,
32\,GB of DRAM, and one NVIDIA GV100 GPU. We ran the experiments in a container 
based on an official Docker image from NVIDIA, which had CUDA 11.4.2, cuDNN 8.0, 
and Ubuntu 20.04 installed.

\stitle{Workloads.}
We evaluated the correctness and efficiency of {\sys} using GPU kernels from 
two GPU benchmarks, Rodinia~\cite{Che2009RodiniaAB} and Parboil~\cite{Stratton2012ParboilAR}, 
and four DL frameworks, TVM~\cite{ApacheTVM} v0.11.0,
PyTorch~\cite{PyTorch} v1.12.1, TensorRT~\cite{TensorRT} v7.2.3.4,
and FasterTransformer~\cite{FasterTransformer} v5.3 (referred to as FT).
For Rodinia, we use kernels and instances from all applications, except for
\texttt{mummergpu}, \texttt{cfd}, \texttt{particlefilter}, \texttt{srad}, 
and \texttt{streamcluster}, as these cannot run on our testbed.
For Parboil, we use kernels and instances of the CUDA-based implementation.
For TVM, PyTorch, and TensorRT, we use kernels and instances for inference tasks 
of five DNN models, ResNet~\cite{He2016DeepRL}, DenseNet~\cite{Huang2017DenselyCC},
VGG~\cite{simonyan2014very}, Inception~\cite{Szegedy2016RethinkingTI},
and MobileNet~\cite{Howard2017MobileNetsEC}.
As for FasterTransformer, we use kernels and instances for an inference task of
GPT-2~\cite{radford2019language}.

\stitle{{\sys} settings.}
In the static analysis phase, we set the maximum number of unroll iterations
to 32 (see $\S$\ref{subsec:strawman}).
As for the preconditions of the kernel parameters utilized for monotonicity
checking (see $\S$\ref{subsec:opt-range}), we establish bounds for each kernel 
parameter by utilizing the minimum and maximum values observed in the evaluated instances.

\subsection{Classification Accuracy}
\label{sec:eval:acc}

\noidentstitle{Methodology.}
We evaluate the accuracy of {\sys} in classifying the idempotency of GPU kernel instances 
(see Table~\ref{tbl:acc}).
First, we employ the idempotency tracing tool (refer to $\S$\ref{subsec:bg-study}) 
to gather the ground truth idempotency (\textbf{Truth}) 
and instances' launch arguments.
We then label the kernel-level idempotency (\textbf{Manual}) as the same way 
mentioned in $\S$\ref{subsec:bg-study}.
Next, we utilize {\sys} to statically analyze the kernel-level idempotency of each GPU kernel (\textbf{Kernel-level}).
Finally, we employ {\sys} to dynamically validate the idempotency of instances using the
launch arguments provided in the trace,
and find incorrect classifications relative to the ground truth (\textbf{Instance-level}).

\stitle{Ground truth.}
We totally evaluate 547 GPU kernels with 18,217 instances (using their public traces), 
where 14,418 instances are idempotent.
All frameworks have both idempotent and non-idempotent GPU kernel instances, 
except for TVM.\footnote{\footnotesize{The TVM community has recently introduced support for 
in-place updates~\cite{TVMInPlaceUpdate}, enabling TVM to generate non-idempotent instances,
though our evaluation encountered only idempotent instances.}}

\begin{figure*}[t]
    \begin{minipage}{1\linewidth}
        \begin{minipage}{.16\linewidth}
            \centering\includegraphics[width=\linewidth]{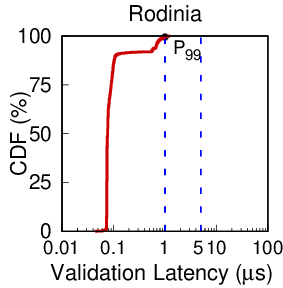} 
        \end{minipage}
        \begin{minipage}{.16\linewidth}
            \centering\includegraphics[width=\linewidth]{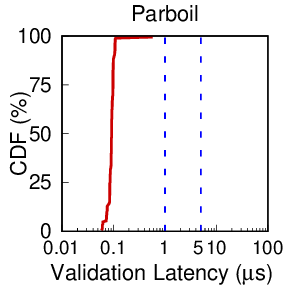} 
        \end{minipage}
        \begin{minipage}{.16\linewidth}
            \centering\includegraphics[width=\linewidth]{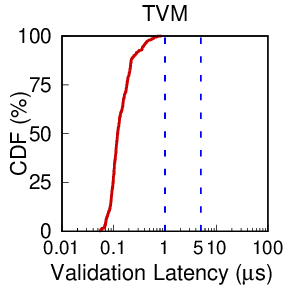}
        \end{minipage}
        \begin{minipage}{.16\linewidth}
            \centering\includegraphics[width=\linewidth]{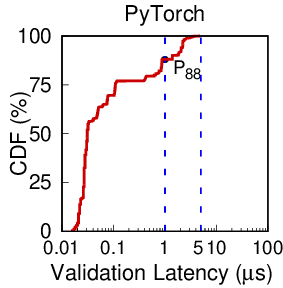}
        \end{minipage}
        \begin{minipage}{.16\linewidth}
            \centering\includegraphics[width=\linewidth]{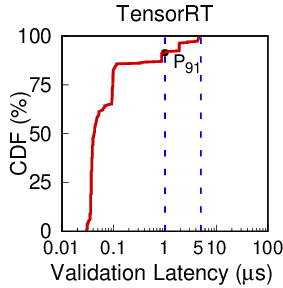}
        \end{minipage}
        \begin{minipage}{.16\linewidth}
            \centering\includegraphics[width=\linewidth]{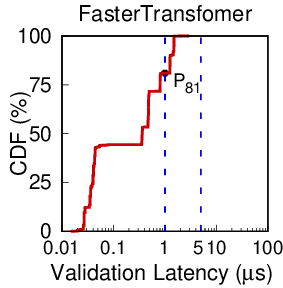}
        \end{minipage}
    \end{minipage}\\[8pt]
    \begin{minipage}{1\linewidth}
    \caption{\emph{\small{The CDF of validation latency for instances in various
    GPU benchmarks and DL frameworks.}}}
    \label{fig:val-latency}
    \end{minipage}  \\[-20pt]
\end{figure*}

\stitle{Kernel-level idempotency.} 
As shown in the group \textbf{Kernel-level} of Table~\ref{tbl:acc},
most kernels (435 out of 547) are classified as {\pidemshort}. 
{\sys} classifies only 16 kernels as idempotent, 
as it requires a kernel to be write-only on GPU memory (e.g., for initialization), 
which is uncommon in practice.
In contrast, 96 kernels are classified as non-idempotent by {\sys}, 
surpassing the manually identified number of 34.
This suggests that some {\pidemshort} kernels are pessimistically classified as non-idempotent.

\stitle{Kernel-level classification breakdown}.
As shown in Table~\ref{tbl:nonidem}, we provide a detailed breakdown of 
why specific kernels are classified as non-idempotent, according to four reasons. 
Specifically, 26 kernels are normally classified as non-idempotent because 
{\sys} can statically identify that they have load and store instructions 
accessing the same symbolic address ({SO}).
There are 51 kernels classified as non-idempotent due to the failure of
symbolic execution, including 8 kernels due to indirect function
calls ({IF}), and 43 kernels due to path explosion ({PE}).
Path explosion is frequently observed with irregular unbounded loops.
The majority of these kernels are matrix multiplication kernels from
cuDNN~\cite{Chetlur2014cuDNNEP} and CUTLASS~\cite{cutlass},
featuring unbounded loops with complex control flows that {\sys} cannot
effectively optimize.
Additionally, 19 kernels are classified as non-idempotent because they have
symbolic addresses containing non-parameter variables, called non-parameter addresses (NA).
Note that not all kernels with non-parameter addresses are
classified as non-idempotent. Only those that have no path conditions to be validated 
at runtime are classified as non-idempotent, as these addresses
will always overlap with others during validation.

\stitle{Instance-level idempotency.}
{\sys} successfully
identifies 11,745 idempotent instances 
(with no false positives), and all non-idempotent instances.
However, 2,673 instances are misclassified as non-idempotent,
resulting in a false-negative rate of 18.54\%.
Instances from TVM exhibit the lowest false-negative rate, at 2.13\%.
This is attributed to the simplicity of the kernels produced by TVM,
which has no unbounded loops that complicate analysis.
On the other hand, other deep learning frameworks have significantly higher
false-negative rates. For example, PyTorch has a false-negative rate
of 35.73\% and TensorRT has a rate of 42.58\%.

\def\falsenegativefootnote{\footnote{Detailed explanations and examples of 
false negatives in {\sys} can be found in Appendix D (see supplementary material).}}

\stitle{Instance-level classification breakdown.}
We further analyze the detailed reasons for false negatives, 
reported in Table~\ref{tbl:fnreason}.
There are 574 false negative instances whose kernels are statically
classified as non-idempotent due to the failure of static analysis (i.e., IF and PE). 
The primary cause of false negatives is non-parameter
address (NA), accounting for 1,723 instances. 
As a reminder, {\sys} adopts a conservative approach by assuming that a non-parameter variable
can be any value, causing non-parameter addresses to always
overlaps with other addresses during validation.
Non-parameter variables can also
exist in path conditions (non-parameter conditions, NC). {\sys} conservatively assumes such 
conditions are always satisfied. This causes the predicted address set 
to exceed the actual one accessed by the instance, resulting in 66 false negatives.
The range-based model in {\sys} causes the remaining 310 false negatives, which
can be correctly classified using our strawman solution.
There are two main types of overestimation. The first type occurs when the concrete
addresses of a symbolic address are discrete but {\sys} considers them as consecutive ranges.
The second type occurs when the range bounds are not tight enough.

\subsection{Validation Latency}
\label{sec:eval:lat}

\noidentstitle{Methodology.}
We next study the validation latency for each instance. 
Only the instances of kernels that are classified as {\pidemshort}
by {\sys} are considered in this experiment, because the validation
function of idempotent and non-idempotent kernels always return true and false,
without any computation.
We execute the validation function of each instance 1,000 times and pick
the longest execution time as the final validation latency for that instance.

\stitle{Validation latency.}
Fig.~\ref{fig:val-latency} illustrates the CDF of the validation latency for
each benchmark and framework. It is worth noting that {\sys} is capable of validating
all instances, across the evaluated benchmarks and frameworks, in under 5\,{\us},
successfully achieving our ideal validation latency design goal (a few microseconds).
Specifically, in Rodinia, PyTorch, TensorRT, and FasterTransformer, 99\%, 88\%, 91\%, and
81\% of the instances, respectively, and all instances in Parboil and TVM, can be validated
within 1\,{\us}.

\begin{figure}[t]
    \begin{minipage}{.48\linewidth}
        \centering\includegraphics[width=\linewidth]{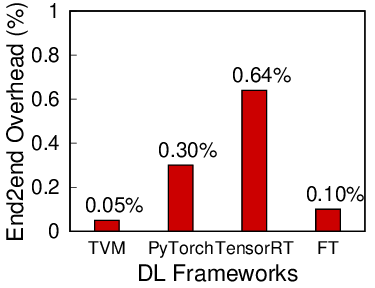} 
    \end{minipage} 
    \begin{minipage}{.48\linewidth}
        \centering\includegraphics[width=\linewidth]{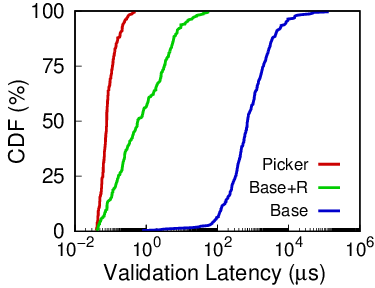} 
    \end{minipage} \\[8pt]
    \begin{minipage}{1\linewidth}
    \caption{\emph{\small{(a) The average overhead on end-to-end performance of DL applications 
	due to dynamic idempotency validation, and (b) The CDF of validation latency for each instance 
	in TVM under various optimization configurations.}}}
    \label{fig:val-overhead}
    \end{minipage} \\[-14pt]
\end{figure}

\stitle{Application overhead.}
We further evaluate the overhead to the application caused by {\sys},
using inference tasks from four DL frameworks.
The results, as shown in Fig.~\ref{fig:val-overhead}~(a), demonstrate that 
the overhead in terms of end-to-end execution time is less than 1\% for all cases.
This is because the end-to-end execution time of tasks is dominated by the
execution time of kernels, and the dynamic validation
of {\sys} is sufficiently fast to overlap with it.

\subsection{Optimization Breakdown}
\label{sec:eval:opt}

\noidentstitle{Methodology.}
To evaluate the efficacy of the optimizations proposed in $\S$\ref{sec:opt}, 
we measure the validation latency for three versions of {\sys}: 
a fully optimized version ({\sys}), a version without range compaction (Base+R), 
and a version without both range compaction and range-based address model (Base). 
All versions enable the compiled execution optimization. 
In this experiment, we specifically use instances from TVM, 
as all its kernels have no unbounded loops, 
allowing for effective analysis with our strawman solution.

\stitle{Validation latency.} 
Fig.~\ref{fig:val-overhead}(b) illustrates the CDF of validation latency 
when enabling different optimizations.
The Base version, which enumerates GPU thread IDs for each symbolic address, 
generates an enormous number of calculated addresses. 
This causes 42\% of instances to have validation latencies over 1\,ms. 
By using range-based address model ($\S$\ref{subsec:opt-range}), 
each symbolic address only requires two calculations for its bounds. 
This significantly reduces the number of calculated addresses, 
bringing the maximum validation latency below 100~{\us}. 
Finally, by applying range compaction ($\S$\ref{subsec:opt3}), 
{\sys} further reduces the number of symbolic addresses, achieving 
a maximum validation latency under 1\,{\us}.

\subsection{Case Studies}
\label{sec:eval:app}

We study two typical cases that improve the performance of idempotence-based systems 
using {\sys} (see \S\ref{subsec:bg-idem-sys}). 

\stitle{Case 1: Checkpoint-based fault tolerant system.}
We implemented a simplified version of Asymmetric Resilience~\cite{Leng2020AsymmetricRE} (AR),
a checkpoint-based GPU fault-tolerant system that leverages the idempotency
to reduce the checkpointing overhead.
Specifically, AR checkpoints the input buffer of every GPU kernel instance to the host memory
before executing the instance. 
If a transient fault (e.g., a bit flip in the registers) occurs during the execution, the system 
is recovered by copying back the saved input data and re-executing the instance.
For idempotent instances, AR avoids the memory checkpointing since it can be recovered
by re-execution.

\begin{figure}[t]
    \begin{minipage}{.48\linewidth}
    \centering\includegraphics[width=\linewidth]{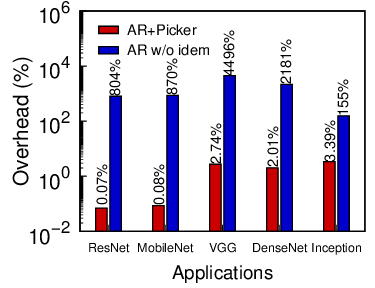} 
    \end{minipage}
    \begin{minipage}{.48\linewidth}
        \centering\includegraphics[width=\linewidth]{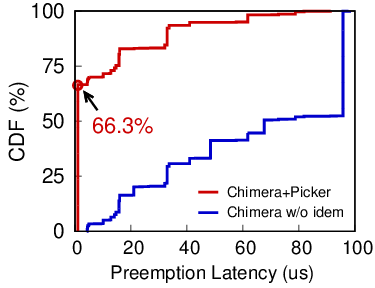} 
    \end{minipage} \\[8pt]
    \begin{minipage}{1\linewidth}
        \caption{\emph{\small{(a) The performance overhead of a checkpoint-based fault 
        tolerant system, and (b) the CDF of preemption latency in a preemptive 
        scheduling system, with and without leveraging the idempotency information
        provided by {\sys}. }}}
    \label{fig:application-case-study}
    \end{minipage} \\[-15pt]
\end{figure}

Fig.~\ref{fig:application-case-study}(a) shows the performance overhead of AR
when executing five TVM-backed DNN inference applications (as mentioned 
in $\S$\ref{sec:eval:setup}).
Without exploiting idempotency (AR w/o idem), the performance overhead 
ranges from 155\% to 4,496\%, making it impractical for real-world applications.
By integrating {\sys}, the performance overhead is reduced to less than 4\% in all cases.
In ResNet and MobileNet, all instances are correctly identified 
as idempotent by {\sys}, eliminating the need for checkpointing and resulting in
negligible overhead (0.08\%) solely from dynamic validation.
The other three applications have instances being 
(mis)classified as non-idempotent (up to 6 instances in Inception), resulting in a 
maximum end-to-end overhead of 3.39\%.

\stitle{Case 2: Preemptive scheduling of inference serving.}
We implemented an extended version of Chimera~\cite{Park2015ChimeraCP}, 
a GPU multitasking system that leverages the idempotency information 
to accelerate preemptive scheduling.
Specifically, if a kernel instance is non-idempotent, 
it is preempted by saving the execution context, generally resulting in 
preemption latency of tens of microseconds~\cite{Lin2016EnablingEP,Tanasi2014EnablingPM}.
In contrast, idempotent instances can be immediately preempted 
by killing them and discarding their execution context,
thereby reducing the preemption latency to less than
1 microsecond~\cite{Park2015ChimeraCP,han2022reef,Lee2021IdempotenceBasedPG}.

Fig.~\ref{fig:application-case-study}(b) shows the CDF of the preemption latency
for GPT-2 inference requests.
Each request consists of 10,000 instances involving 50 kernels~\cite{FasterTransformer}. 
{\sys} identifies 5,803 instances as idempotent.
The latency to kill a running instance is assumed to be 1\,{\us}, 
following prior work~\cite{Park2015ChimeraCP,han2022reef,Lee2021IdempotenceBasedPG}.
Without exploiting idempotency (Chimera w/o idem), 
preemption latency ranges from 4\,{\us} to 98\,{\us},
depending on the context size of preempted instances, with an average of 64.3\,{\us}.
By integrating {\sys}, 66.3\% of the preemptions are completed within 1\,{\us}, 
reducing the average latency by 84.2\% to 10.2\,{\us}.

\section{Related Work}

\noidentstitle{Systems using idempotence.}
The idempotence property has been widely used in many systems, such as
serverless computing~\cite{Zhang2020FaulttolerantAT,Sreekanti2020AFS,Jia2021BokiSS,
Ding2023AutomatedV, Qi2023HalfmoonLF,Jangda2019FormalFO}, 
persistency model~\cite{Lin2019ExploringMP,DBLP:conf/iiswc/YudhaKZS20},
storage and distributed systems~\cite{Kim2006ExploitingRI,Sigurbjarnarson2016PushButtonVO,Helland2012IdempotenceIN,ramalingam2013fault,Zhuang2023EXOFLOW}
and intermittent computing~\cite{Surbatovich2019IODI,Woude2016IntermittentCW}.
Flux~\cite{Ding2023AutomatedV} is one of these systems that are most related to {\sys}, 
which can automatically verify the idempotency for a group of concurrent stateful 
serverless functions. 
The key difference between serverless functions and GPU kernels is the way they 
access state. Serverless functions access external databases using the table name and 
the record key, with the table name usually being a constant value, thereby making 
it possible to statically determine the idempotency. On the other hand, GPU kernels
access GPU memory using pointers, with the base address typically being a parameter of
the kernel and unknown before invocation. Therefore, {\sys} chooses to dynamically
validate the idempotency.

\stitle{GPU program analysis.}
One of the key techniques behind {\sys} is the automatic program analysis of GPU kernels.
There are many existing researches on GPU program analysis for different purposes, such as
bug detection~\cite{Li2012ParametricFA,Chiang2013FormalAO,Li2010ScalableSV},
race detection~\cite{Collingbourne2014SymbolicCO,Betts2012GPUVerifyAV,Betts2015TheDA,Bardsley2014EngineeringAS, Li2014PracticalSR, Boyer2008AutomatedDA,Kamath2021iGUARDIA},
memory coalescing detection~\cite{Alur2017GPUDranoDU,Alur2021StaticDO},
test generation~\cite{Li2012GKLEECV, Leung2012VerifyingGK}
and security~\cite{Mai2023HONEYCOMB}.
Honeycomb~\cite{Mai2023HONEYCOMB} is one of these systems that are most related to {\sys}, which
is a software-based TEE for GPU applications. Honeycomb leverages static program analysis to
validate the security of GPU kernels. 
Both {\sys} and Honeycomb focus on validating the memory addresses accessed by an instance.
However, there are two major technical distinctions.
First, Honeycomb solely validates whether all memory accesses made by an instance fall
within the specified memory regions. {\sys} validates whether the read and write accesses overlap.
Second, Honeycomb utilizes a polyhedral model to represent the memory addresses that
an instruction may access. This approach is efficient and requires a small code base.
On the contrary, {\sys} leverages monotonicity, which is more accurate but
requires the evolvement of SMT solver.

\section{Conclusion}

This paper presents {\sys},
the first system that dynamically validates the instance-level idempotency on the GPU.
Evaluation shows the efficacy and efficiency of {\sys}.


\small{
\bibliographystyle{plain} 
\bibliography{gpu-idempotence}
}

\end{document}